\documentclass[twocolumn,showpacs,preprintnumbers,prl]{revtex4}
\usepackage{amssymb}
\usepackage{amsmath}
\usepackage{graphicx}
\usepackage{dcolumn}
\usepackage{bm}

\begin{document}

\title{Quantum Dissonance Is Rejected in an Overlap Measurement Scheme}
\author{Chang-shui Yu$^{1}$}
\email{quaninformation@sina.com; ycs@dlut.edu.cn}
\author{Jun Zhang$^{1}$}
\author{Heng Fan$^{2}$}
\affiliation{$^{1}$School of Physics and Optoelectronic Technology, Dalian University of
Technology, Dalian 116024, P. R. China\\
$^{2}$Beijing National Laboratory for Condensed Matter Physics, Institute of
Physics, Chinese Academy of Sciences, Beijing 100190, China}
\date{\today }

\begin{abstract}
The overlap measurement scheme accomplishes to evaluate the overlap of two input quantum states by only measuring an introduced auxiliary qubit, irrespective of the complexity of the two input states. We find a counterintuitive phenomenon that no quantum dissonance can be found, even though the auxiliary qubit might be entangled, classically correlated or even  uncorrelated with the two input states based on different types of input states. In principle, this provides an
opposite but supplementary example to the remarkable algorithm of the deterministic quantum
computation with one qubit in which no entanglement is present. 
Finally, 
we consider a simple overlap measurement model to demonstrate the 
continuous change (including potential sudden death of quantum discord) with the input states from entangled to product states  by only adjusting some simple initial parameters.

\end{abstract}

\maketitle

\textit{Introduction.}---Quantum discord [1-3] beyond quantum entanglement [4] can effectively grasp the role of
quantumness of correlations and distinguish them from the
classical correlations [5].  It has been
attracting increasing interests in various areas such as the evolution in
quantum dynamical systems [6-10], the operational interpretation [11-17],
the quantification [18-20] and so on. However, even 
though
strong evidences, 
in theory [21] and experiment [22], have shown that some QIPT displayed the quantum advantage without any quantum
entanglement, the role of various correlation in quantum information processing tasks (QIPT) is only restricted to several remarkable cases [23,24].
The most remarkable is the algorithm of deterministic
quantum computation with one qubit (DQC1)[23,24]. It shows that there is no quantum entanglement present in the process and  quantum dissonance [18,23]-----the quantum correlation without
entanglement might be related to the speedup of DQC1, which changes our 
 intuitive understanding about the role of entanglement in the (QIPT). Therefore, when one deals with some QIPTs, 
 one will naturally consider which one between  quantum discord and entanglement plays the key role in the QIPT.
 With such an emphasis in mind,  one could easily think that (1) if a QIPT with quantum advantage
is accomplished without quantum entanglement, one would expect quantum
dissonance. In addition, since quantum correlation `includes' quantum
entanglement due to the potential presence in separable states
[6,25,26], one could also think that (2) if the correlations with the quantum state in a  QIPT are changed from quantum entanglement to classical correlations, whilst the quantum advantage is always present, there should exist some states for which quantum dissonance
would be present. We will show that neither is generally  true.

In this Letter, we study a surprising overlap measurement scheme (OMS)
[27,28] which accomplishes to evaluate the overlap of two states by only
measuring a single auxiliary qubit irrespective of the complexity of the
measured states. We shows that the OMS for some
input states can be successfully implemented only with the classical correlation and even surprisingly without any correlations between the auxiliary qubit and the two measured quantum states. As is unlike what we could expect either, at any rate, one can not find the quantum dissonance in the OMS, even though there could exist quantum entanglement.
This provides an opposite example to the model of DQC1
[23], here we show that only the quantum dissonance is unnecessary. It could also let us reconsider
 what is the source of the speedup in the QIPT including DQC1, as is suspected by Ref. [20].
As a
simple example, we consider the overlap measurement of the states of two
two-level particles undergoing a depolarizing channel, respectively. In this
simple model, we show that the OMS is accomplished with the continuous change from the absence to the
presence of the correlations by only changing some simple parameters. 
Besides, one can also find that quantum discord accompanied by entanglement can  suddenly die with the properly adjusted  parameter of the input states.

\textit{The overlap measurement scheme.}-To begin with, we would like to
briefly introduce the OMS, which can be shown by the quantum circuit given
in Fig.1. Two quantum states $\rho _{1}$ and $\rho _{2}$ as the measured
states are input into the quantum circuit. The third qubit is introduced as
the auxiliary one which is prepared initially in the state $\left\vert
0\right\rangle $.  A Hadamard operation $H$ is performed on the auxiliary
qubit and then a CSWAP gate (controlled-swap) $C_{swap}$ is operated on the
three particles with the auxiliary qubit as the control qubit and the other two as the controlled ones, where the Hadamard operation is given by $H\left\vert
0\right\rangle =\frac{1}{\sqrt{2}}\left( \left\vert 0\right\rangle
+\left\vert 1\right\rangle \right) $ and the CSWAP gate is given by $\left\{
\begin{array}{c}
C_{swap}\left\vert 0\right\rangle \left\vert \Phi \right\rangle \left\vert
\Psi \right\rangle =\left\vert 0\right\rangle \left\vert \Phi \right\rangle
\left\vert \Psi \right\rangle  \\
C_{swap}\left\vert 1\right\rangle \left\vert \Phi \right\rangle \left\vert
\Psi \right\rangle =\left\vert 1\right\rangle \left\vert \Psi \right\rangle
\left\vert \Phi \right\rangle
\end{array}%
\right. $ for any states $\left\vert \Psi \right\rangle $ and $\left\vert
\Phi \right\rangle $ with $\left\vert i\right\rangle,i=0,1$ as the auxiliary qubit. A measurement device in the $\sigma _{x}$ basis is
placed at the end of the auxiliary qubit in order to read out the final
state of the qubit. The probability $p(\pm )$ corresponding to the outcomes $%
\left\vert \pm \right\rangle =\frac{1}{\sqrt{2}}\left( \left\vert
0\right\rangle \pm \left\vert 1\right\rangle \right) $ is given by%
\begin{equation*}
p_{\pm }=\frac{1}{2}\left( 1\pm \text{Tr}\rho _{1}\rho _{2}\right) .
\end{equation*}%
Thus the overlap of $\rho _{1}$ and $\rho _{2}$ can be obtained by a single
parameter $p_{\pm }$, irrespective of the complexity of the measured states.
This scheme is quite useful, because it can be used to measure the geometric
distance of two quantum states and some quantum entanglement witness [27].
In particular, if the input states are the two copies of the given state $%
\varrho $, $p_{\pm }$ will be directly related to the purity Tr$\varrho ^{2}$
which can be employed to evaluate the Renyi entropy [27], bipartite
entanglement of pure states [28] and the quantumness of a single quantum
state [29,30]. Next we will study what kind of correlations are necessary for this powerful scheme.

\textit{Correlations in the scheme.}---In order to find out the roles of
various correlations in the OMS, we have to repeat the scheme in
mathematics. Consider two input particles prepared in the states $\rho _{1}$
and $\rho _{2}$, respectively, the final state after thewhole operations shown in Fig. 1 can be
given by%
\begin{eqnarray}
\rho _{a12} &=&C_{swap}\left[ \left( H\left\vert 0\right\rangle
_{a}\left\langle 0\right\vert H^{\dag }\right) \otimes \rho _{1}\otimes \rho
_{2}\right] C_{swap}^{\dag }  \notag \\
&=&\frac{1}{2}\left(
\begin{array}{cc}
\rho _{1}\otimes \rho _{2} & \left( \rho _{1}\otimes \rho _{2}\right)
S \\
S\left( \rho _{1}\otimes \rho _{2}\right)  & \rho _{2}\otimes \rho
_{1}%
\end{array}%
\right) ,
\end{eqnarray}%
where the subscript ``$a$'' denotes the auxiliary qubit and $S$ is the swap gate
 defined by $S\left\vert\Psi\right\rangle\left\vert\Phi\right\rangle=\left\vert\Phi\right\rangle\left\vert\Psi\right\rangle$  for any states $\left\vert \Psi \right\rangle $ and $\left\vert
\Phi \right\rangle $. Based on Eq. (1), we
will have the following useful theorems.
\begin{figure}[tbp]
\includegraphics[width=0.8\columnwidth,bb=0 320 600 550]{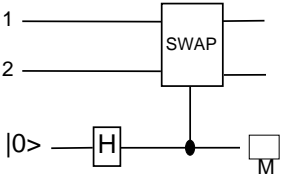}
\caption{The quantum circuits of the OMS. Two states
can be input from the port 1 and 2. The third state $\left\vert
0\right\rangle $ as an auxiliary qubit goes through the Hadamard gate and
then undergoes the controlled-swap gate simultaneously with the rest two
input particles and is finally measured on  $\protect\sigma _{x}$
by the device $M$. }
\label{1}
\end{figure}

\textbf{Theorem 1.} If $\rho _{1}\neq \rho _{2}$, the auxiliary qubit will
be entangled with the two input particles $1$ and $2$, i.e., $\rho _{a\left(
12\right) }$ is entangled in the OMS.

\textbf{Proof.} In order to show there exists quantum entanglement between the
auxiliary qubit and the two input particles, we will employ the negativity
as separability criterion [31]. That is, if the partial transpose of the
density matrix $\rho _{a\left( 12\right) }$ is not positive, our theorem
will hold. From Eq. (1), one can find that the partial transpose of the
density matrix $\rho _{a\left( 12\right) }$ can be given by%
\begin{equation}
\rho _{a12}^{T_{12}}=\frac{1}{2}\left(
\begin{array}{cc}
\rho _{1}\otimes \rho _{2} & S\left( \rho _{1}\otimes \rho
_{2}\right)  \\
\left( \rho _{1}\otimes \rho _{2}\right) S & \rho _{2}\otimes \rho
_{1}%
\end{array}%
\right) ^{\ast },
\end{equation}%
where the superscript $T_{12}$ means the transpose on the subsystems $1$ and
$2$. Let $\rho _{1}^{\ast }=\sum\limits_{i=1}^{r_{1}}\lambda _{i}\left\vert
\varphi _{i}\right\rangle \left\langle \varphi _{i}\right\vert $ and $\rho
_{2}^{\ast }=\sum\limits_{i=1}^{r_{2}}\sigma _{i}\left\vert \psi
_{i}\right\rangle \left\langle \psi _{i}\right\vert $ be the eigenvalue
decompositions of $\rho _{1}^{\ast }$ and $\rho _{2}^{\ast }$ with $\lambda
_{i}$ and $\sigma _{i}$ arranged in the decreasing order, where $r_{i}$ is
the rank of $\rho _{i}$. Next we will prove this theorem in four cases. 

(i)
Assume there exists an $n<\min \left\{ r_{1},r_{2}\right\} $ such that $%
\left\langle \varphi _{i}\right\vert \left. \psi _{i}\right\rangle =1$ for $%
i\leq n$ (if there is no such $n$ that satisfies this condition, $n=0$), then we have
\begin{equation}
\left\langle \varphi _{i}\right\vert \left. \psi _{j}\right\rangle
=\left\langle \varphi _{i}\right\vert \left. \varphi _{j}\right\rangle =0%
\text{ for }j>n,i\leq n.
\end{equation}%
In this case, if we construct the vector $\left\vert x\right\rangle =\frac{1%
}{\sqrt{2}}\left[
\begin{array}{c}
-\left\vert \psi _{n+1}\right\rangle \left\vert \varphi _{n+1}\right\rangle
\\
\left\vert \varphi _{n+1}\right\rangle \left\vert \psi _{n+1}\right\rangle
\end{array}%
\right] $, then
\begin{gather}
\left\langle x\right\vert \rho _{a12}^{T_{12}}\left\vert x\right\rangle
=\left\langle \psi _{n+1}\right\vert \rho _{1}\left\vert \psi
_{n+1}\right\rangle \left\langle \varphi _{n+1}\right\vert \rho
_{2}\left\vert \varphi _{n+1}\right\rangle /2  \notag \\
-\left\langle \varphi _{n+1}\right\vert \rho _{1}\left\vert \varphi
_{n+1}\right\rangle \left\langle \psi _{n+1}\right\vert \rho _{2}\left\vert
\psi _{n+1}\right\rangle /2  \notag \\
=-\lambda _{n+1}\sigma _{n+1}/2+\left\langle \psi _{n+1}\right\vert \rho
_{1}\left\vert \psi _{n+1}\right\rangle \left\langle \varphi
_{n+1}\right\vert \rho _{2}\left\vert \varphi _{n+1}\right\rangle /2.
\end{gather}%
From Eq. (3), we have $\left\langle \psi _{n+1}\right\vert \rho
_{1}\left\vert \psi _{n+1}\right\rangle <\lambda _{n+1}$ and $\left\langle
\varphi _{n+1}\right\vert \rho _{2}\left\vert \varphi _{n+1}\right\rangle
<\sigma _{n+1}$, so\ Eq. (4) will arrive at%
\begin{equation}
\left\langle x\right\vert \rho _{a12}^{T_{12}}\left\vert x\right\rangle <0%
\text{.}
\end{equation}%

(ii) If $n=$ $r_{1}<r_{2}$, we have $\left\langle \varphi _{i}\right\vert
\left. \psi _{i}\right\rangle =1$ and $\left\langle \varphi _{i}\right\vert
\left. \psi _{j}\right\rangle =0$ for $i\leq n$ and $j>n$. Thus we set $%
\left\vert x^{\prime }\right\rangle =\frac{1}{\sqrt{2}}\left[
\begin{array}{c}
-\left\vert \psi _{n+1}\right\rangle \left\vert \varphi _{n}\right\rangle
\\
\left\vert \varphi _{n}\right\rangle \left\vert \psi _{n+1}\right\rangle
\end{array}%
\right] $, so

\begin{gather}
\left\langle x^{\prime }\right\vert \rho _{a12}^{T_{12}}\left\vert x^{\prime
}\right\rangle =\left\langle \psi _{n+1}\right\vert \rho _{1}\left\vert \psi
_{n+1}\right\rangle \left\langle \varphi _{n}\right\vert \rho _{2}\left\vert
\varphi _{n}\right\rangle /2  \notag \\
-\left\langle \varphi _{n}\right\vert \rho _{1}\left\vert \varphi
_{n}\right\rangle \left\langle \psi _{n+1}\right\vert \rho _{2}\left\vert
\psi _{n+1}\right\rangle /2  \notag \\
=-\lambda _{n}\sigma _{n+1}/2<0.
\end{gather}%

(iii) If $n=$ $r_{2}<r_{1}$, we have $\left\langle \varphi _{j}\right\vert
\left. \psi _{j}\right\rangle =1$ and $\left\langle \varphi _{i}\right\vert
\left. \psi _{j}\right\rangle =0$ for $i>n$ and  $j\leq n$. Thus we set $%
\left\vert x^{^{\prime \prime }}\right\rangle =\frac{1}{\sqrt{2}}\left[
\begin{array}{c}
-\left\vert \psi _{n}\right\rangle \left\vert \varphi _{n+1}\right\rangle
\\
\left\vert \varphi _{n+1}\right\rangle \left\vert \psi _{n}\right\rangle
\end{array}%
\right] $. Based on the similar calculation as Eq. (6), we have

\begin{equation}
\left\langle x^{\prime \prime }\right\vert \rho _{a12}^{T_{12}}\left\vert
x^{\prime \prime }\right\rangle =-\lambda _{n+1}\sigma _{n}/2<0.
\end{equation}%

(iv) If $n=$ $r_{1}=r_{2}$, we have $\left\langle \varphi _{i}\right\vert
\left. \psi _{j}\right\rangle =\delta _{ij}$ for $i,j\leq n$. Due to $\rho
_{1}\neq \rho _{2}$, it is impossible that the vectors $[\lambda
_{1},\lambda _{2},\cdots ,\lambda _{n}]^{T}$ and $[\sigma _{1},\sigma
_{2},\cdots ,\sigma _{n}]^{T}$ made up of the eigenvalues are linearly
dependent. So there must exist integers $k$ and $l$ such that $\frac{\lambda
_{k}}{\lambda _{l}}>\frac{\sigma _{k}}{\sigma _{l}}$, i.e. $\lambda
_{k}\sigma _{l}>\lambda _{l}\sigma _{k}$. Now we set $\left\vert x^{\prime
\prime \prime }\right\rangle =\frac{1}{\sqrt{2}}\left[
\begin{array}{c}
-\left\vert \psi _{l}\right\rangle \left\vert \varphi _{k}\right\rangle  \\
\left\vert \varphi _{k}\right\rangle \left\vert \psi _{l}\right\rangle
\end{array}%
\right] $, we can obtain
\begin{gather}
\left\langle x^{\prime \prime \prime }\right\vert \rho
_{a12}^{T_{12}}\left\vert x^{\prime \prime \prime }\right\rangle
=\left\langle \psi _{l}\right\vert \rho _{1}\left\vert \psi
_{l}\right\rangle \left\langle \varphi _{k}\right\vert \rho _{2}\left\vert
\varphi _{k}\right\rangle /2  \notag \\
-\left\langle \varphi _{k}\right\vert \rho _{1}\left\vert \varphi
_{k}\right\rangle \left\langle \psi _{l}\right\vert \rho _{2}\left\vert \psi
_{l}\right\rangle /2  \notag \\
=\left( -\lambda _{k}\sigma _{l}+\lambda _{l}\sigma _{k}\right) /2<0.
\end{gather}%
Hence, Eqs. (5-8) show us that $\rho _{a12}^{T_{12}}$ is not positive for
any $\rho _{1}\neq \rho _{2}$, which implies that $\rho _{a\left( 12\right) }
$ is entangled.\hfill$\blacksquare$

\textbf{Theorem 2.} If $\rho _{1}=\rho _{2}$ are mixed states, there will
only exist classical correlation between the auxiliary qubit and the input
particles $1$ and $2$.

\textbf{Proof.} Since $\rho _{1}=\rho _{2}$, we can set the eigenvalue decompositions
of the two density matrices to be $\rho _{i}=\sum\limits_{i=1}^{r}\lambda
_{i}\left\vert \psi _{i}\right\rangle \left\langle \psi _{i}\right\vert $
with $r$ denoting the rank of the matrices. Substitute the eigenvalue
decomposition into Eq. (1), one will have%
\begin{gather}
\rho _{a\left( 12\right) }=\frac{1}{2}\sum \lambda _{i}\lambda _{j}\left(
\left\vert 0\right\rangle _{a}\left\langle 0\right\vert \right. \otimes
\left\vert \psi _{i}\psi _{j}\right\rangle _{12}\left\langle \psi _{i}\psi
_{j}\right\vert   \notag \\
+\left\vert 0\right\rangle _{a}\left\langle 1\right\vert \otimes \left\vert
\psi _{i}\psi _{j}\right\rangle _{12}\left\langle \psi _{j}\psi
_{i}\right\vert +\left\vert 1\right\rangle _{a}\left\langle 0\right\vert
\otimes \left\vert \psi _{j}\psi _{i}\right\rangle _{12}\left\langle \psi
_{i}\psi _{j}\right\vert   \notag \\
+\left\vert 1\right\rangle _{a}\left\langle 1\right\vert \otimes \left.
\left\vert \psi _{j}\psi _{i}\right\rangle _{12}\left\langle \psi _{j}\psi
_{i}\right\vert \right)   \notag \\
=\frac{1}{2}\left\vert +\right\rangle _{a}\left\langle +\right\vert \otimes
\sum \lambda _{i}\lambda _{j}\left\vert \Psi _{ij}^{+}\right\rangle
_{12}\left\langle \Psi _{ij}^{+}\right\vert   \notag \\
+\frac{1}{2}\left\vert -\right\rangle _{a}\left\langle -\right\vert \otimes
\sum \lambda _{i}\lambda _{j}\left\vert \Psi _{ij}^{-}\right\rangle
_{12}\left\langle \Psi _{ij}^{-}\right\vert ,
\end{gather}%
where $\left\vert \Psi _{ij}^{\pm }\right\rangle _{12}=\frac{1}{\sqrt{2}}%
\left( \left\vert \psi _{i}\psi _{j}\right\rangle _{12}\pm \left\vert \psi
_{j}\psi _{i}\right\rangle _{12}\right) $. Thus one can easily find that $%
\left\langle \Psi _{ij}^{m}\right. \left\vert \Psi _{lk}^{n}\right\rangle
=\delta _{il}\delta _{jk}\delta _{mn}$ and $\left\langle \Psi
_{ij}^{+}\right. \left\vert \Psi _{ii}^{+}\right\rangle =2\delta _{ij}$. Eq.
(9) can also be considered as one eigenvalue decomposition of $\rho _{a12}$.
A direct observation of Eq. (9) shows that not only quantum entanglement but
also any quantum correlation is not present between the auxiliary qubit and
the two input particles. The reduced density matrices of the auxiliary qubit
and the two input particles can be calculated as%
\begin{equation}
\rho _{a}=p_{+}\left\vert +\right\rangle _{a}\left\langle +\right\vert
+p_{-}\left\vert -\right\rangle _{a}\left\langle -\right\vert ,
\end{equation}%
and%
\begin{equation}
\rho _{12}=\frac{1}{2}\sum \lambda _{i}\lambda _{j}\left( \left\vert \Psi
_{ij}^{+}\right\rangle _{12}\left\langle \Psi _{ij}^{+}\right\vert
+\left\vert \Psi _{ij}^{-}\right\rangle _{12}\left\langle \Psi
_{ij}^{-}\right\vert \right) .
\end{equation}%
From Eqs (9-11), the total correlation based on the mutual information [2]
can be given by%
\begin{eqnarray}
I(\rho _{a(12)}) &=&S(\rho _{a})+S(\rho _{12})-S(\rho _{a(12)})  \notag \\
&=&-p_{+}\log _{2}(p_{+})-p_{-}\log _{2}(p_{-})  \notag \\
&\neq &0.
\end{eqnarray}%
Thus, Eqs. (9) and (12) show that $\rho _{a(12)}$ has zero quantum
correlation and nonzero total correlation, which implies that only classical
correlation is present in the OMS if the two input mixed states have the
same density matrix.\hfill$\blacksquare$

\textbf{Corollary 3. }If $\rho _{1}=\rho _{2}=\left\vert \phi \right\rangle
\left\langle \phi \right\vert $ (i.e., they are the same pure state in
Theorem 2), there will not be any correlation between the auxiliary qubit
and the two input particles.

\textbf{Proof.} Substitute $\rho _{1}=\rho _{2}=\left\vert \phi \right\rangle
\left\langle \phi \right\vert $ into Eq. (9), one can easily have
\begin{equation}
\rho _{a\left( 12\right) }=\left\vert +\right\rangle _{a}\left\langle
+\right\vert \otimes \left\vert \phi \right\rangle \left\langle \phi
\right\vert \otimes \left\vert \phi \right\rangle \left\langle \phi
\right\vert ,
\end{equation}%
which shows no correlation in $\rho _{a\left( 12\right) }$.\hfill$\blacksquare$

For the Corollary 3, one could think that the purity measurement of a pure
state is quite trivial, since the measured quantum state is pure and hence
has purity one. However, considering the scheme of the purity measurement in
a practical scenario, it is not necessary for us to know whether the
measured quantum state is pure or not. In other words, if the measurement
outcome is 1, we will draw the conclusion that the input two states are the
same two pure states and no correlation is ever present in the process. Thus
the surprising conclusion is that the purity measurement of a pure state
requires no correlation. From Theorem 2, one can find another interesting
thing that the purity measurement of a mixed state based on this scheme only
requires classical correlation instead of quantum correlation. As a summary,
we can safely say that the purity measurement does not need any quantum
correlation, which might be unlike what we had expected. 
In addition, our theorems also provide us an intuitional picture of the continuous change of 
correlations from the entanglement to product states in the OMS, which is given in
Fig. 2. Fig. 2 shows  that in the OMS, there does not exist the set
of the states with quantum dissonance. The states change
directly from the set of entangled states to the set of classically
correlated states, and the quantum correlation and quantum entanglement of the
 final state are present and vanish simultaneously with the change of the initial states.
 Thus we can conjecture that sudden death of quantum discord on the parameters of the states could be present, which is verified by our latter example.
\begin{figure}[tbp]
\includegraphics[width=0.6\columnwidth,bb=0 250 600 600]{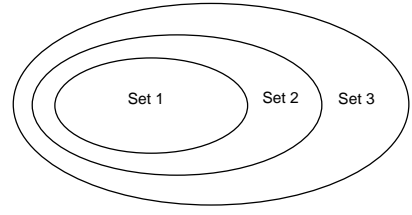}
\caption{The illustration of the relationship between various state sets in
the overlap measurement scheme. Sets $1$, $2$, and $3$ correspond to the
sets of quantum entangled states, classically correlated states and product
states, respectively.}
\label{1}
\end{figure}

\textit{A simple example.}-As an example, we would like to consider a toy
model to demonstrate the presence of the different correlations in the
measurement process. The model is sketched in Fig. 3. Suppose we have two
two-level input particles. Both are initially prepared in the state $%
\left\vert 0\right\rangle _{i}$ where $i=1,2$ corresponds to the different
input particles and we use $\left\vert 0\right\rangle $ and $\left\vert
1\right\rangle $ to distinguish the two orthogonal levels. Let the two
particles undergo two depolarizing quantum channels $\$_{i}$ which, for any
a density matrix $\chi $, are given by [32]
\begin{equation}
\varrho _{i}=\$_{i}\left( \chi \right) =a_{i}\chi +\frac{1-a_{i}}{2}\mathbf{1%
},
\end{equation}%
where $\mathbf{1}$ is the identity and we assume that $a_{i}\in \lbrack -1,1]
$, is separately controlled. In order to measure the overlap of $\varrho _{1}
$ and $\varrho _{2}$, we need let the two states into the measurement
devices corresponding to the quantum circuit given in Fig. 1.
Mathematically, we need to substitute $\varrho _{i}$ into Eq. (1) and to
obtain the corresponding $\rho _{a(12)}^{\prime }(a_{1},a_{2})$. In this
way, we will have the total correlation of $\rho _{a(12)}^{\prime
}(a_{1},a_{2})$ as%
\begin{equation}
I\left( \rho _{a(12)}^{\prime }(a_{1},a_{2})\right) =S(\rho_a^\prime)+S(\rho_{12}^\prime)-S(\rho_{a(12)}^\prime)\end{equation}%
with $\rho_{a}^\prime=Tr_{12}\rho_{a(12)}^\prime,  \rho_{12}^\prime=Tr_{a}\rho_{a(12)}^\prime$. In particular, if $a_{1}=a_{2}=\pm 1$, we will find that $I\left(
\rho _{a(12)}^{\prime }(a_{1},a_{2})\right) =0$, which means that there is
no correlation in the system of interests. This case corresponds to the
purity measurement of a pure state, which is consistent with our Corollary 3. 
Based on a simple calculation, we can also find that the negativity can be
given by
\begin{equation}
N\left( \rho _{a(12)}^{\prime }(a_{1},a_{2})\right) =\left\vert
a_{1}-a_{2}\right\vert .
\end{equation}%
It is obvious that the system of the auxiliary qubit and the rest will
disentangle for $a_{1}=a_{2}=a$ which means that the two input states are
the same, that is, our scheme corresponds to the purity measurement. In this
case, $\rho _{a(12)}^{\prime }(a_{1},a_{2})$ can be written as the form of
Eq. (9), with $\lambda _{1}=\frac{1+a}{2}$, $\lambda _{2}=\frac{1-a}{2}$ and
$\left\vert \Psi _{ij}^{\pm }\right\rangle =\frac{1}{\sqrt{2}}\left(
\left\vert ij\right\rangle \pm \left\vert ji\right\rangle \right) $, $i,j=0,1
$. Thus, we show that there is no quantum correlation in the system by this
example. The illustration of the change of correlations is given in Fig. 3.

In general, it is hard to show the sudden death of quantum discord in  decoherence due to the exponential  
decay of the entries of the considered density matrix [33]. Here, we assume the parameters of the initial states exponentially depend
on time $t$ as $a_1=e^{-\Gamma_1t}$ and $a_2=e^{-\Gamma_2t}$ with $\Gamma_1=2\Gamma_2=10$ for $t<=0.2$ and $\Gamma_1=\Gamma_2=10$ for the rest. The initial values 
are given by $a_{10}=1$ and  $a_{20}=\frac{1}{e}$.
Thus we have found that quantum discord and negativity simultaneously die at $t=0.2$, which is shown by the \textit{black} solid line in the lower layer of Fig. 3.
\begin{figure}[tbp]
\includegraphics[width=1\columnwidth]{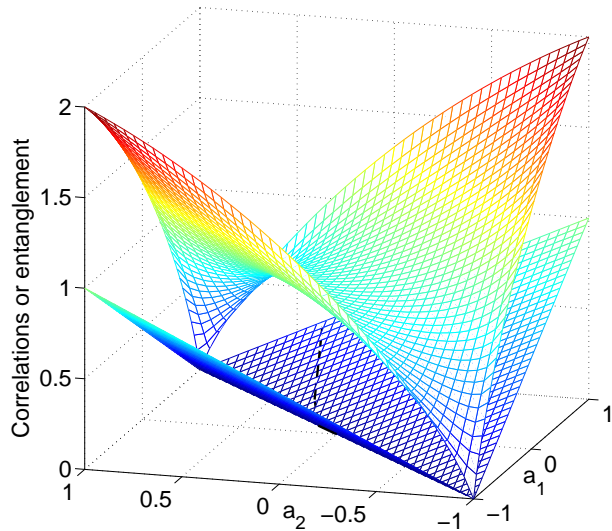}
\caption{(Color online,Dimensionless)The total correlation and negativity $%
vs.$ $a_{1}$ and $a_{2}$. There are two layers in the figure. The upper
layer denotes the total correlation and the lower one means the negativity. $%
a_{1}=a_{2}=\pm 1$ corresponds to no correlation, $a_{1}=a_{2}\neq \pm 1$
corresponds to no entanglement, meanwhile the total correlation can be
understood as classical correlation, and the other cases of $a_1,a_2$c orresponds to the
presence of quantum entanglement. The black solid line in the lower layer shows one path that leads to the sudden death on initial parameters of quantum discord and entanglement.}
\label{1}
\end{figure}

\textit{Conclusions and discussion.}--- We have studied the various
correlations in the OMS. It is shown that the OMS could require the presence of quantum entanglement, classical correlations and even no correlations based on the different types of input states. However, at any rate, we can not find the existence of the quantum dissonance in the OMS.  This provides
a supplementary example to the model studied in Refs. [23,24] where only quantum
entanglement is not necessary. It also let us be in doubt that quantum discord is the source of the speedup 
in some QIPT such as DQC1. In addition, we find that 
quantum correlation and quantum entanglement are present and vanish simultaneously on the initial parameters due to the absence of quantum dissonance. 
Thus the parameters of the initial states could lead to the sudden death of quantum correlation.

 We also give a simple
model as an example to demonstrate the presence and the absence of various
correlations in the scheme by only adjusting the parameters $a_{i}$.
Experimentally, the scheme shown in FIG.1 can be realized in some standard 
quantum information processing systems such as nuclear magnetic resonance system,
optical system, ion-trap system, superconducting qubit system etc.

Finally, we would like to raise some interesting questions. How can we find some schemes that demonstrate the absence of correlation along the
line from quantum entanglement, quantum dissonance, classical correlation to
the product states? Since there is no quantum correlation in some QIPT with quantum advantage,  if we still contribute the advantage to quantum correlation,
it maybe need  to introduce some other measures of the quantumness of correlations, or challengingly, to consider other candidate as  the source of the speedup of these QIPTs. 
We hope that the present work will add new viewpoint to our understanding of
the power of quantumness.

\textit{Acknowledgements.}-This work was supported by the National Natural
Science Foundation of China, under Grant No. 11175033
and `973' program No. 2010CB922904 and the Fundamental Research Funds of the Central Universities, under Grant No. DUT12LK42.


\begin{thebibliography}{99}
\bibitem{[29]} Since for any $n\times n$ density matrix excluding $\varrho =%
\mathbf{1}/n$, there exists quantumness witness based on Theorem 2 in Ref.
[30], one can measure the quantuness based on the distance defined by $%
d(\rho ,\mathbf{1}/n)=$Tr$\left\vert \rho -\mathbf{1}/n\right\vert ^{2}=$Tr$%
\rho ^{2}-\frac{1}{n}$.
\bibitem{[33]} The sudden death is considered in the sense of the changed parameters of initial states, which is different from the case of decoherence.
\end{thebibliography}
\end{document}